\documentclass[letterpaper,preprintnumbers,prd,twocolumn,nofootinbib,nobibnotes,showpacs]{revtex4}
\usepackage{amsfonts}
\usepackage{mathrsfs}
\usepackage{epsfig}
\usepackage{graphicx}%
\usepackage{dcolumn}
\usepackage{amsmath}
\usepackage{color}
\usepackage{color}
\makeatletter
\def\btt#1{\texttt{\@backslashchar#1}}%
\DeclareRobustCommand\bblash{\btt{\@backslashchar}}%
\makeatother
\begin{document}
\title{Quasinormal modes of static and spherically symmetric black holes with the derivative coupling}
\author{Shuang Yu}\email{yushuang@nao.cas.cn}
\affiliation{ Key Laboratory of Computational Astrophysics, National Astronomical Observatories, Chinese
Academy of Sciences, Beijing 100101, China}
\affiliation{School of Astronomy and Space Sciences, University of Chinese Academy of Sciences,
No.19A, Yuquan Road, Beijing 100049, China}
\author{Changjun Gao}\email{gaocj@bao.ac.cn}
\affiliation{ Key Laboratory of Computational Astrophysics, National Astronomical Observatories, Chinese
Academy of Sciences, Beijing 100101, China}
\affiliation{School of Astronomy and Space Sciences, University of Chinese Academy of Sciences,
No.19A, Yuquan Road, Beijing 100049, China}
\date{\today}
%%%%%%%%%%%%%%%%%%%%%%%%%%%%%%%%%%%%%%%%%%%%%%%%%%%%%%%%%%%%%%%%%%%%%%%%%%
\begin{abstract}
We investigate the quasinormal modes of a class of static and spherically symmetric black holes with the derivative coupling.
The derivative coupling has rarely been paid attention to the study of black hole quasinormal modes. Specifically, we study the effect of derivative coupling on the quasinormal modes for four kinds of black holes. They are Reissner-Nordstrom black holes, Bardeen black holes, noncommunicative geometry inspired  black holes and dilaton black holes. These black holes are not the solutions of vacuum Einstein equations which guarantees the effect of derivative coupling is not trivial. We find the influence of derivative coupling on the quasinormal modes roughly mimics the overtone numbers. In other words,  there is a qualitative similarity in the trend of quasinormal modes frequencies due to increasing either the coupling constant and the overtone number.
\end{abstract}

%\keywords{}

\maketitle

%%%%%%%%%%%%%%%%%%%%%%%%%%%%%%%%%%%%%%%%%%%%%%%%%%%%%%%%%%%%%%%%%%%%%%%%%%
\section{Introduction}
Recent detections of gravitational waves at LIGO and VIRGO \cite{ligo,virgo} mark the beginning of the astronomical era of gravitational waves.
The dominating contribution to gravitational waves when a black hole is perturbed comes from the quasinormal modes. This is a long period of damping proper oscillations whose frequencies only depend on the parameters of the black hole. Thus it is called the ``character sound'' of black holes.
Making perturbations to a black hole can be performed in two ways. One is by
adding matter fields of different spins to the black hole spacetime and the other is by perturbing
the black hole metric itself. Either way, the black hole undergo damped oscillations at the intermediate
stage with complex frequencies. The real part of the frequency
describes the oscillation rate and the imaginary part describes the {damped} rate. These oscillations are called quasi-normal
modes (QNMs).

Black hole QNMs  are in the first place studied in the gravitational and electromagnetic perturbations around black hole spacetimes, late-time evolution of fields in the black hole spacetime and numerical simulations of stellar collapse \cite{Kokkotas:1999,Nollert:1999,Regge:1957,Zerilli:1970,Vishveshwara2:1970,Vishveshwara:1970}.
And then it is found that black hole QNMs can be associated with the restoration of thermal equilibrium for the perturbed state in ADS/CFT \cite{adscft}.
Thirdly, it is revealed there is a connection between QNMs and spacetime geometries with the Choptuik scaling \cite{conn}.
Finally, QNMs have also become relevant to quantum gravity for some different reasons. It arises from
Hod's proposal of describing quantum properties of black holes from their classical oscillation spectrum \cite{hod}.
So far, black hole QNMs have been extensively studied from different perspectives. For an incomplete list see \cite {quasi} and references therein.

On the other hand, in order to explain the acceleration of the Universe, many extensions of general relativity
have been proposed. In particular, a lot of work has been devoted to Horndeski theories \cite{Horn} which is derived
by Horndeski forty years ago and independently re-derived by \cite{deff} recently. Horndeski theories are the most general scalar-tensor theories with second order equation of motion.
Thus they are free of ghosts. It is found that the Horndeski theories play important role in both cosmological and astrophysical
scenarios, for example, inflationary cosmology\cite{inflation}, dark energy cosmology \cite{dark energy}
neutron stars \cite{n-stars} and other astrophysical objects \cite{astro-ob}.

One of the many interesting features of Horndeski
theories is the coupling between the derivative of
the scalar field and the Einstein tensor.
In the aspect of cosmology, this term leads to cosmic speed-up
without the need of any scalar potential. This is firstly pointed out in \cite{first}.
The studies on this coupling have attracted much interest
in both inflationary \cite{inflationary} and late time cosmology \cite{latetime}.
However, rare attention of this coupling has been paid to the black hole quasinormal modes so far.
Some studies in this aspect are as follows.
{Ref.\cite{chenjingprd:2010} studied the QNMs and the dynamical evolution of a scalar field with the derivative coupling for charged black hole.}
Ref.\cite{Zhang:2018} studied the QNMs of a massive scalar field with the derivative coupling for a regular black hole.
Ref. \cite{Yao:2011} investigated the QNMs of a scalar perturbation with this derivative coupling in the warped ADS black hole spacetime. {Ref. \cite{fontana:2018} investigated the QNMs and dynamical evolution of the scalar perturbation with this derivative coupling in de Sitter spacetimes. Finally, Ref. \cite{konoplya:2018} investigated the long-lived quasinormal modes and instability problem of a massive non-minimally coupled scalar field in Reissner-Nordstrom spacetime.}

Thus the goal of this paper is to investigate the quasinormal
modes of a massive scalar field with derivative coupling in the most general, static and spherically symmetric black hole spacetimes. The analysis of the QNMs in our calculations may provide a hint for the identification of derivative coupling in the scalar field from the future observations of gravitational waves. To this end, we shall start from the Klein-Gordon equation with the derivative coupling and derive the effective potential in the background of most general, static and spherically symmetric black hole spacetimes. The effective potential plays a key role in the calculation of QNMs. Then four black hole spacetimes are considered. They are Reissner-Nordstr\"om black holes (RN),  Bardeen black holes \cite{bardeen:1968}, noncommunicative geometry inspired (NGI) black holes \cite{Nicolini:2006} and dilaton black holes \cite{Gibbons:1988, Garfinkle:1991}. Why do we consider these four black holes? We notice that in order to illustrate the effect of derivative coupling, $G_{\mu\nu}$ must not be vanishing. So the solutions of vacuum Einstein equations, such as the \emph{Schwarzschild }solution, the Kerr
solution and their de Sitter extension, would present us trivial results. Therefore we consider above four black holes with non-vanishing Einstein tensors.

The paper is organized as follows. In section II, we derive the effective potential in general, static and spherically symmetric black hole spacetimes and describe the third-order WKB method. It is perhaps the most popular method for the calculation of black hole QNMs which is devised by Schutz, Will and Iyer \cite{WKB:1987,WKB:1988,Iyer:1985}. In section III, we study the effect of derivative coupling on the effective potential. The expression of effective potential for Bardeen black holes, NGI black holes and dilaton black holes are rather lengthy. Therefore, we do not explicitly write them here and only make their plotting. There are two phases for RN, Bardeen and NGI black holes. One is for the extreme black holes and the other is for the non-extreme case. The extreme black holes have only one horizon and the non-extreme have two horizons. So the effect of derivative coupling on these two scenarios are discussed. In section IV, we tabulate and make an analysis on the corresponding QNMs for four kinds of black holes.
Finally, section IV gives the conclusion and discussion. Throughout the paper, the system of units is $G=c=\hbar=1$ and the metric signature is $(-,\ +,\ +,\ +)$.

\section{effective potential and the WKB method}
In this section, we derive the effective potential for the scalar field with derivative coupling in the background of general static and spherically symmetric black hole spacetimes. The corresponding metric is given by
\begin{eqnarray}
d{{s}^{2}}=-f\left(r\right)d{{t}^{2}}+f{{\left(r\right)}^{-1}}d{{r}^{2}}+R{{\left(r\right)}^{2}}d{{\Omega }^{2}},
\end{eqnarray}
where $d{{\Omega }^{2}}=d{{\theta }^{2}}+{{\sin }^{2}}\theta d{{\varphi }^{2}}$ is the line element for a unit sphere. We consider the scalar field with derivative coupling which has the equation of motion as follows
\begin{eqnarray}
\square \Phi +\beta {{G}_{\mu \nu }}{{\nabla }^{\mu }}{{\nabla }^{\nu }}\Phi+{{m}^{2}}\Phi =0\;.
\end{eqnarray}
Here $m$ is the mass of scalar particle, $\beta$ is the coupling  constant which has the dimension of square of length. In the absence of $\beta$ term, it restores to the
usual Klein-Gordon equation. The $\beta$ term has attracted much interest
in both inflationary \cite{inflationary} and late time cosmology \cite{latetime}. However, rare attention of this term  has been paid to the black hole quasinormal modes. Making separation of the field $\Phi(t,r,\theta ,\varphi)$ as follows
\begin{eqnarray}
\Phi (t,r,\theta ,\varphi )= {e^{ - i\omega t}}{Y_{lm}}(\theta ,\varphi )K(r),
\end{eqnarray}
we obtain the radial equation for the scalar perturbation
\begin{eqnarray}
A{K}''+B{K}'+\left({{\omega }^{2}}C-\tilde{V}\right)K=0,
\end{eqnarray}
where ${K}''=\frac{{{\partial }^{2}}K}{\partial {{r}^{2}}}$, ${K}'=\frac{\partial K}{\partial r}$, and $A,B,C,\tilde{V}$ are defined by
\begin{eqnarray}
A\equiv{{g}^{11}}+\beta {{G}^{11}}\;,
\end{eqnarray}
\begin{eqnarray}
B\equiv\frac{1}{{R{{\left(r\right)}^2}}}\left[{{R}}{\left(r\right)^2}\left(g^{11} + \beta G^{11}\right)\right]'\;,
\end{eqnarray}
\begin{eqnarray}
C\equiv-\left({g^{00}}+\beta {G^{00}}\right)\;,
\end{eqnarray}
\begin{eqnarray}
\tilde{V}\equiv{m^2} + l\left(l + 1\right)\left({g^{22}} + \beta {{{G}}^{22}}\right)\;.
\end{eqnarray}
Here $l=0,\ 1,\ 2,\cdot\cdot\cdot$ and $'$ denotes the derivative with respect to $r$.  $Y_{lm}$ is the spherical harmonics. In order to analyse the frequencies of quasinormal modes, we should transform the radial equation into the standard form
\begin{eqnarray}\label{standard}
&&F_{,r_{\ast}r_{\ast}}+\left[\omega^2-{V}\left(r\right)\right]F=0\;,
\end{eqnarray}
where $F$, $r_{\ast}$ and $V$ are the new radial function,  the new radial coordinate and the effective potential, respectively. ``$,$'' denotes the derivative with respect to $r_{\ast}$. To this end, we introduce a function $E$ and replace the radial function $K$ with $F$ defined by
\begin{eqnarray}
F\equiv CEK,
\end{eqnarray}
 then Eq.(4) becomes
\begin{eqnarray}
\left\{\frac{{{d}^{2}}}{{{d}^{2}}{{r}_{*}}}+{{\omega }^{2}}-\frac{\tilde{V}}{C}+\left[\left(A+B\right)E\frac{{{d}^{2}}}{{{d}^{2}}r}\left(\frac{1}{CE}\right)\right]\right\}F=0\;,
\end{eqnarray}
provided that
\begin{eqnarray}
\frac{1}{2}\left(\frac{AE}{CE}\right)'=\frac{BE}{CE}+2AE\left(\frac{1}{CE}\right)'.
\end{eqnarray}
This equation determines the expression of function $E$.
$r_{\ast}$ is defined by
\begin{eqnarray}
 {{{r}_{*}}}\equiv\int\sqrt{\frac{C}{A}}{dr}\;.
\end{eqnarray}
Then we obtain the effective potential
\begin{eqnarray}
{V}&=&\frac{\tilde{V}}{C}-E\left[A\left(\frac{1}{CE}\right)''+B\left(\frac{1}{CE}\right)'\right]\;.
\end{eqnarray}
Eliminating $E$ by using Eq.~(12), we obtain
\begin{eqnarray}
{V}&=&- A{ \zeta ^\prime } +  \zeta \left\{ AC\left[\left(\frac{1}{C}\right)' -  \zeta \right] - B\right\}  + \frac{\tilde{V}}{C}\;,
\end{eqnarray}
where
\begin{eqnarray}
\zeta\equiv\frac{1}{{2A}}\left[\frac{1}{2}\left(\frac{A}{C}\right)'-\frac{B}{C}\right]\;.
\end{eqnarray}
{This result is useful since it can be applied to arbitrary static spherically symmetric black hole spacetimes. We note that for the RN black hole, the potential was firstly derived by Chen and Jing \cite{chenjingprd:2010}.} In the next sections, we shall study the quasinormal modes of Reissner-Nordstrom black holes, dilaton black holes, Bardeen black holes and noncommunicative geometry inspired black holes, respectively.

To calculate the QNMs of black holes, various numerical methods have been proposed in literature.
They are Mashhoon method\cite{mashhoon:1984}, Chandrasekhar-Detweiler and shooting methods\cite{chand},
WKB method\cite{WKB:1987,WKB:1988,Iyer:1985}, continued fraction method \cite{Leaver:1985},
P\"oschl-Teller approximation\cite{Ferrari:1984}, and phase integral method \cite{Andersson:1992,Andersson2:1992}.
In this paper, we shall evaluate the quasinormal modes of black holes by using the third-order WKB method.
It is perhaps the most popular method. This method has been used extensively in evaluating quasinormal frequencies of various black holes. For an incomplete list see \cite {quasi} and references therein. {We note that even higher order WKB method \cite{RAK:2003,JM:2017} could produce much more accurate results than the third-order method. But for simplicity in calculations, we shall adopt the third-order WKB method}.

The frequency of QNMs given by the third-order WKB method \cite{WKB:1987,WKB:1988} takes the form
\begin{eqnarray}
{\omega ^2} = \left({V_0} + \sqrt { - 2{V_0}^{\prime \prime }} \Lambda \right) - i\left(n + \frac{1}{2}\right)\sqrt { - 2{V_0}^{\prime \prime }} \left(1 + \Omega \right)\;,
\end{eqnarray}
{where $\Lambda$ and $\Omega$ are determined by the overtone number $n$ and the derivatives of the effective potential at the peak. The bulky expressions of $\Lambda$ and $\Omega$ can be found in \cite{WKB:1987,WKB:1988}}. $n=0,\ 1,\ 2, \cdot\cdot\cdot$ are the overtone numbers and $r_p$ is the position for the peak of effective potential. It is pointed that \cite{car:04} the accuracy of the WKB method depends on the multipolar
number $l$ and the overtone number $n$. The WKB approach is consistent with the numerical method very well provided that  $l>n$. Therefore we shall
present the quasinormal frequencies of scalar perturbation for $n=0$ and $l=1,2,3$, respectively. In the next sections, we analyse the effect of derivative coupling on the potential and calculate the quasinormal modes of Reissner-Nordstrome black holes,
dilaton black holes, Bardeen black holes and noncommunicative geometry inspired black holes, respectively {while assume the mass $m$ of scalar is zero.}

\section{potential of the four kinds of black holes}
\subsection{RN black holes}
For RN black holes, we have the metric functions $f(r)$ and $R(r)$ as follows
\begin{eqnarray}
f\left(r\right)=1-\frac{2M}{r}+\frac{{{Q}^{2}}}{{{r}^{2}}}\;,
\end{eqnarray}
\begin{eqnarray}
R\left(r\right)=r\;,
\end{eqnarray}
where $M$ and $Q$ are the mass and charge of the black hole, respectively. Using Eqs.~(15-16) and Eqs.~(5-8), we obtain the corresponding effective potential
\begin{eqnarray}
V&=&\frac{{{r^2}-2Mr+{Q^2}}}{{{r^6}{{\left({r^4}+\beta {Q^2}\right)}^2}}}\left\{{Q^4}{\beta^2}\left[4{Q^2} + {r^2}\left(2-{l^2}-l\right)-6Mr\right]\right.\nonumber\\&& \left.
+\beta{r^4}{Q^2}\left(6{Q^2}+6{r^2}+{m^2}{r^4}-12rM\right)\right.\nonumber\\&& \left.
+{r^8}\left(-2{Q^2}+{r^4}{m^2}+{r^2}{l^2}+2Mr+{r^2}l\right)\right\}\;.
\end{eqnarray}
Then we are ready to compute the QNMs by using Eq.~(17). Observing this potential, we find the coupling constant $\beta$ can not be negative. Otherwise, the potential would be divergent at $r=\sqrt[4]{-\beta Q^2}$.  {The divergence of the potential would lead to the presence of ghosts. This has been verified by Germani and Kehagias at the level of the Horndeski models \cite{germani:2010,germani:2011}.} Therefore we shall consider positive $\beta$ in the following. In general, there are two horizons in the RN spacetime. They are the outer event horizon ${r_+}=M+\sqrt {{M^2}-{Q^2}}$  and the inner Cauchy horizon ${r_-}=M-\sqrt{{M^2}-{Q^2}}$, respectively. When $M=Q$, the two horizons coincide and the situation is called extreme RN spacetime. In the next subsections, let's consider the two situations, respectively.

\subsubsection{Extreme RN black holes}

Setting $M=1$ and $Q=1$, we get an extreme RN black hole. In Fig.~1, we plot the effect of coupling constants $\beta$ on the shape of the potential.
We choose the other parameters as follows: $l=1$ and $m=0$. The event horizon locates at $r_{\ast}=-\infty$ for the tortoise coordinate. From the figure we see, with the increasing of coupling constants $\beta$, the height of the potential is reduced. Furthermore, the potential become negative in some regions.

{The presence of the potential well implies the instability of the black hole against scalar perturbations \cite{chenjingprd:2010}. Actually, Chen and Jing \cite{chenjingprd:2010} have shown that the instability always occurs when the coupling constant $\beta$ is larger than a certain threshold value
$\beta_c$ provided that $l>0$.  When $0\leq\beta\leq\beta_c$, the potential is always positive and the black hole is stable against scalar perturbations. For $l=0$, there does not exist such a threshold value and the scalar field is always stable for arbitrary coupling constant. We will see in the following the potential well also appears for the Bardeen black holes and the NGI black holes.}

 Since the horizon locates at $r_{\ast}=-\infty$, we conclude it is only in the vicinity of event horizon that the potential is significantly affected by the derivative coupling.

\begin{figure}[h]
\includegraphics[width=9cm]{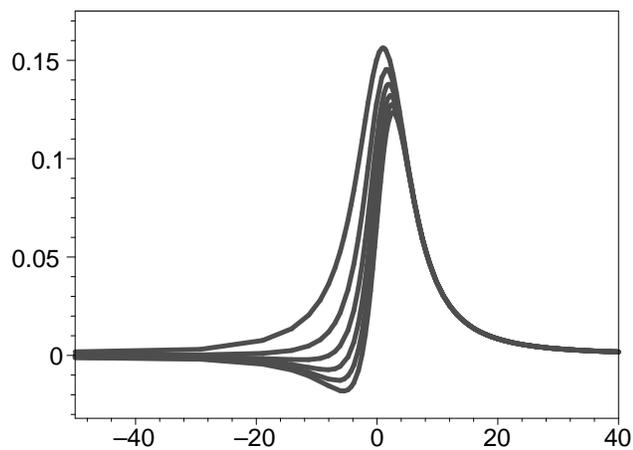}
\caption{The effective potential $V(r_{\ast})$ of extreme RN black holes as a function of the tortoise coordinate $r_\ast$ assuming $M=1, Q=1, m=0, l=1$ for four different cases $\beta=0,1,2,3,4,5$, from up to down, respectively.}\label{pot1}
\end{figure}

\subsubsection{Non-extreme RN black holes}
When we set $\text{M}=1$, $\text{Q}=0.95$, we would get the non-extreme RN black hole. In FIG.2, we plot the variation of the effective potential with coupling constants.
From the figure we see the height of the potential is on the decrease with the increasing of coupling constants. This is the same as the extreme RN black holes.
The potential is asymptotically vanishing when $r_*=-\infty$ and $r_*=+\infty$. In order to give a clear demonstration of how $\beta$ influences
the evolution of potential, we have put $\text{Q}=0.95$. When $\text{Q}$ is much small, we find the potential is very insensitive to the variation of the coupling constant.
The reason for this is that the term $\beta G_{\mu\nu}\nabla^{\mu}\phi\nabla^{\nu}\phi$ is proportional to $\text{Q}^2$.

\begin{figure}[h]
\includegraphics[width=9cm]{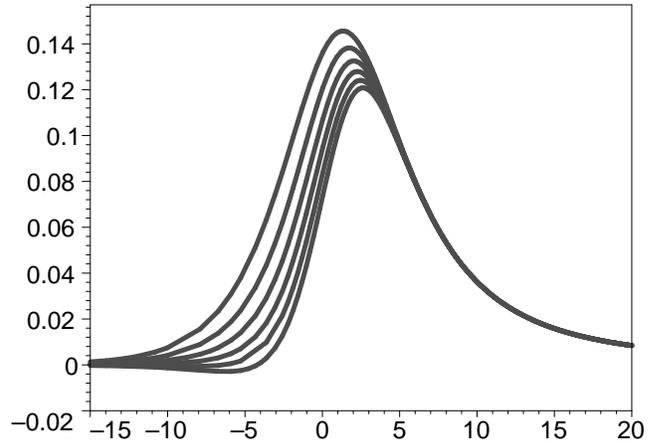}
\caption{The effective potential $V(r_{\ast})$ of non-extreme RN black holes as a function of the tortoise coordinate $r_\ast$ assuming $M=1, Q=0.95, m=0, l=1$ for four different cases $\beta=0,1,2,3,4,5$, from up to down, respectively.}\label{pot2}
\end{figure}

\subsection{Bardeen  black hole}

Bardeen black hole \cite{bardeen:1968} is a regular black hole in the absence of a central singularity.  E. Ayon-Beato and A. {Garcia} show that the Bardeen black hole is an exact solution of the Einstein equations with a magnetic monopole \cite{bardeen:1998}. The functions $f(r)$ and $R(r)$ in the metric are
\begin{eqnarray}
f\left(r\right)&=&1-\frac{2Mr^2}{\left(r^2+Q^2\right)^{3/2}}\;,\nonumber\\
R(r)&=&r\;,
\end{eqnarray}
where $M$ and $Q$ are the mass and charge of black hole, respectively. When $r$ is sufficiently small, we obtain a de Sitter core for this spacetime. On the other hand, when $r$ is large enough, we obtain the Schwarzschild solution. Therefore, this is a regular spacetime. The horizon is determined by
\begin{eqnarray}
1-\frac{2Mr^2}{\left(r^2+Q^2\right)^{3/2}}=0\;,
\end{eqnarray}
which gives the radii of two horizons:
\begin{eqnarray}
r_{+}&=&\left[\chi+2\sqrt{\eta}\cos{\left(\frac{1}{3}\arccos{\frac{\xi}{\eta^{3/2}}}\right)}\right]^{1/2}\;,\nonumber\\
r_{-}&=&\left[\chi+2\sqrt{\eta}\cos{\left(\frac{1}{3}\arccos{\frac{\xi}{\eta^{3/2}}}-\frac{2\pi}{3}\right)}\right]^{1/2}\;,
\end{eqnarray}
where
\begin{eqnarray}
\chi&\equiv&\frac{4}{3}M^2-Q^2\;,\ \ \
\eta\equiv\frac{16}{9}M^4-\frac{8}{3}Q^2M^2\;,\nonumber\\
\xi&\equiv&2Q^4M^2+\frac{64}{27}M^6-\frac{16}{3}Q^2M^4\;,
\end{eqnarray}
provided that
\begin{eqnarray}
Q^2<\frac{16}{27}M^2\;.
\end{eqnarray}
Here $r_{+}$ is the outer event horizon and $r_{-}$ is inner horizon. If $Q=0$, we have $r_{+}=2M$ and $r_{-}=0$. This is for the schwarzschild solution. If
\begin{eqnarray}
Q^2=\frac{16}{27}M^2\;,
\end{eqnarray}
the two horizons coincide and we obtain the extreme Bardeen black hole. On the other hand, if
\begin{eqnarray}
Q^2>\frac{16}{27}M^2\;,
\end{eqnarray}
there would be no horizon in this spacetime. So in the next subsections, let's consider the extreme and non-extreme cases, respectively.
\subsubsection{Extreme Bardeen black holes}
Compared with RN black holes, the expressions of effective potential for Bardeen black holes, noncommutative geometry inspired Schwarzschild black holes,
dilaton black holes, are rather lengthy. Therefore, we do not explicitly write them here and only make their plotting. By setting $M=1$, $Q=4\sqrt{3}/9\simeq0.7698$, $m=0$, $l=1$, we can plot the evolution of the effective potential with $\beta$ in FIG.3.

\begin{figure}[h]
\includegraphics[width=9cm]{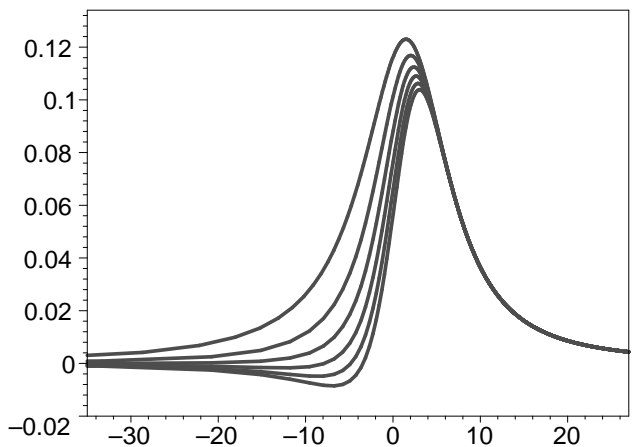}
\caption{The effective potential $V(r_{\ast})$ of extreme Bardeen black holes as a function of the tortoise coordinate $r_\ast$ assuming $M=1, Q=4\sqrt{3}/9, m=0, l=1$ for four different cases $\beta=0,1,2,3,4,5$, from up to down, respectively.}\label{pot3}
\end{figure}

\subsubsection{Non-extreme Bardeen black holes}
By setting $Q=0.76$, $M=1$, $m=0$, $l=1$, we can plot the evolution of the effective potential with $\beta$ in FIG.4. Same as the non-extreme RN black holes,
the potential is insensitive to $\beta$  for sufficiently small $Q$ because the coupling term $\beta G_{\mu\nu}\nabla^{\mu}\phi\nabla^{\nu}\phi\propto \text{Q}^2$.
In order to give a clear demonstration of how $\beta$ influences the evolution of potential, we have put $\text{Q}=0.76$ which is very close to the critical value $0.7698$.

\begin{figure}[h]
\includegraphics[width=8cm]{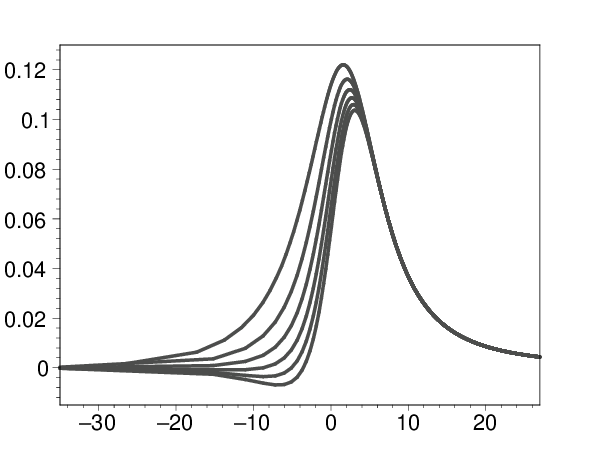}
\caption{The effective potential $V(r_{\ast})$ of non-extreme Bardeen black holes as a function of the tortoise coordinate $r_\ast$ assuming $M=1, Q=0.76, m=0, l=1$ for four different cases $\beta=0,1,2,3,4,5$, from up to down, respectively.}
\end{figure}

\subsection{Noncommunicative geometry inspired Schwarzschild black hole}
 NGI ({Noncommunicative geometry inspired) black hole is derived by Nicolini et.al.\cite{Nicolini:2006}. The functions $f(r)$ and $R(r)$ are found to be
\begin{eqnarray}
f(r)&=&1-\frac{2M}{r}\textrm{erf}\left(\frac{r}{2\sqrt{\theta}}\right)+\frac{2M}{\sqrt{\pi\theta}}e^{-\frac{r^2}{4\theta}}\;,\nonumber\\
R(r)&=&r\;,
\end{eqnarray}
where $M$ is the mass diffused through a region of size $\sqrt{\theta}$, and $\theta$ is a constant with dimension of length squared. When $r\rightarrow \infty$ (or $\theta\rightarrow 0$), we have asymptotically $f=1-2M/r$. It is restored to the Schwarzschild solution.  On the other hand, when $r\rightarrow 0$ (or $\theta\rightarrow \infty$), we have $f=1-s_0 r^2$ ($s_0$ is a constant). It is the de Sitter solution in the core. Thus the NGI Schwarzschild black hole is a regular black hole solution without central singularity but the horizons remain. Furthermore, there are usually two horizons in this spacetime. If
\begin{eqnarray}
\theta<0.2758M^2\;,
\end{eqnarray}
we have two horizons. If
 \begin{eqnarray}
\theta=0.2758M^2\;,
\end{eqnarray}
the two horizons coincide and we have one single horizon. This is the extreme NGI black hole. On the other hand, if
 \begin{eqnarray}
\theta>0.2758M^2\;,
\end{eqnarray}
there would be no horizon in this spacetime.
By putting $M=1$ and $\theta=0.2758$, we can plot the effective potential of extreme NGI black holes for $l=1$ and $m=0$ with different $\beta$ in Fig.~5.

\begin{figure}[h]
\includegraphics[width=8cm]{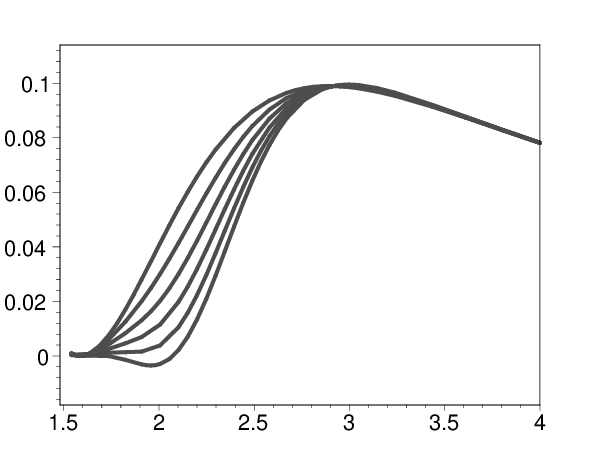}
\caption{The effective potential $V(r)$ of extreme NGI black holes as a function of coordinate $r$ assuming $M=1, \theta=0.2758, m=0, l=1$ for six different cases $\beta=0,1,2,3,4,5$, from up to down, respectively.}
\end{figure}

Similarly, by putting $M=1$ and $\theta=0.24$, we can plot the effective potential for the non-extreme NGI black holes in Fig.~6.

\begin{figure}[h]
\includegraphics[width=8cm]{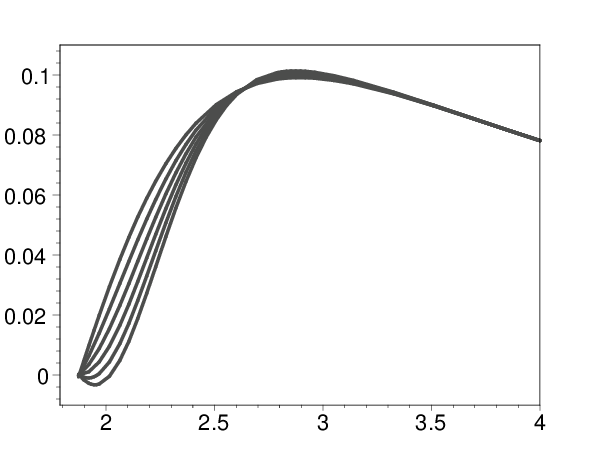}
\caption{The effective potential $V(r)$ of non-extreme NGI black holes as a function of coordinate $r$ assuming $M=1, \theta=0.24, m=0, l=1$ for six different cases $\beta=0,1,2,3,4,5$, from up to down, respectively.}
\end{figure}

We emphasise that in order to show the effects of different $\beta$ on the curves clearly, we use the physical coordinate $r$ instead of the tortoise coordinate
 $r_{\ast}$ in the figure. The outer event horizon locates at $r=1.6$ and $r=1.85$ for extreme NGI and non-extreme NGI, respectively. In tortoise coordinate system,
they all correspond to $r_{\ast}=-\infty$. As shown in the figures,
the potential is asymptotically vanishing at the event horizons.

\subsection{Dilaton black hole}
The RN black holes, Bardeen black holes and NGI black holes bear the same feature of $R(r)=r$ in the line element. In this subsection, we consider
the dilaton black holes which have the different form of $R(r)$. Dilaton black hole was found by Gibbons and Maeda \cite{Gibbons:1988} as well as Garfinkle et. al\cite{Garfinkle:1991} in string theory. {The study on QNMs of dilaton black holes without the derivative coupling can be found in Refs.\cite{dilaton:qnm}.}

The functions $f$ and $R$ are given by
\begin{eqnarray}
f(r)&=&1- \frac{{2M}}{r}\;,\nonumber\\
R(r)&=& \sqrt {r(r - 2D)}\;,
\end{eqnarray}
where $M$ and $D$ are the mass and charge of the black hole. This spacetime has only one horizon. By setting $D=0.5$ and $M=1$, we plot the evolution of the
effective potential with $\beta$ in FIG.7. Same as the NGI black holes, in order to show the effects of different $\beta$ on the curves clearly, we replace the tortoise coordinate $r_{\ast}$ with
physical coordinate $r$. The outer event horizon locates at $r=2$. It corresponds to $r_{\ast}=-\infty$ in tortoise coordinate system.

Observing the evolution of effective potential for the four kinds of black holes, we find that they are all reduced with the
increasing of coupling constants $\beta$ except for the NGI black holes. They are all asymptotically vanishing at the spatial infinity and the event horizon. It is only
for the extreme black holes that the influence of coupling constant $\beta$ on the effective potential is significant. For the non-extreme black hole,
the effect of coupling constant is negligible. { These observations are consistent with the findings of Ref.\cite{chenjingprd:2010} }

\begin{figure}[h]
\includegraphics[width=8cm]{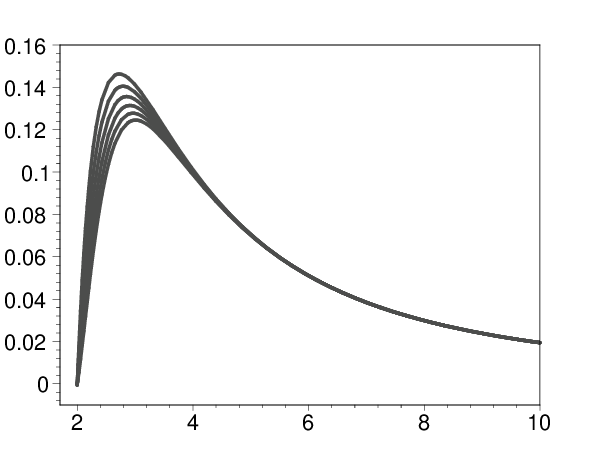}
\caption{The effective potential $V(r)$ of dilaton black holes as a function of coordinate $r$ assuming $M=1, D=0.5, m=0, l=1$ for six different cases $\beta=0,1,2,3,4,5$, from up to down, respectively.}
\end{figure}

\section{quasinormal modes}
{Before the calculation of black hole QNMs, we emphasize that the WKB method mentioned previously does not catch possible instabilities which have been indicated by the effective potentials. In fact, Chen and Jing \cite{chenjingprd:2010} have found that the scalar field is unstable in the background of RN spacetime for some range of parameters. This was also confirmed recently by Konoplya et. al \cite{konoplya:2018}. Specifically, they find that the
effective potential of the extreme RN black hole is always positive-definite within the following range:
\begin{eqnarray}
0\leq\beta\leq 1\;,
\end{eqnarray}
which guarantees stability as shown in Fig.~(1) and Fig.~(2). On the contrary, when
\begin{eqnarray}
\beta>1\;,
\end{eqnarray}
a negative gap emerges in the potential and, according to \cite{taka:2009}, leads to eikonal instability. We find above constraints on $\beta$ also apply to
Bardeen black holes and NGI black holes which can be clearly seen from Fig.~(3-6). So in the next, we shall present the QNMs for $\beta\leq 1$}.

From TABLE.I to TABLE.VII, we list the QNMs of four kinds of black holes with the increasing of coupling constants $\beta$ for $l=0, 1, 2, 3$. {In order to exhibit the difference of frequency with the variation of coupling constant or overtone number, at least three digits should be kept. So we let all the frequencies having four significant digits.} The data show
that the real and imaginary part of QNMs have different trend of evolutions. Namely, the real part is on the decrease and the imaginary part is on the increase
 with the increasing of coupling constant $\beta$. The increasing of imaginary part means the decay of scalar perturbation
becomes faster and faster with the increasing of $\beta$. On the other hand, the decreasing of real part means the oscillating of scalar perturbation
becomes slower and slower with the increasing of $\beta$. This is closely related to the fact that the height of potential is greatly reduced with
the increasing of coupling constant.

From TABLE.VIII to TABLE.XIV we present the QNMs for different $l$ and different $n$ with $\beta=1$. Because it is under the condition $l>n$ that
the WKB approach is applicable  \cite{car:04}, we compute the quasinormal frequencies of scalar perturbation for $l>n$, respectively.
{However, if the precision required is not high, one may also calculate the spherical mode $l=n=0$ using the WKB method, for example, as done in Ref.\cite{chenjing:2010}.} {Similar to }the  scenario of $\beta$, the real part of QNMs is on the decrease and the imaginary part is on the increase
 with the increasing of coupling constant $n$. The increasing of imaginary part means the decay of scalar perturbation
becomes faster and faster with the increasing of $n$. The decreasing of real part means the oscillating of scalar perturbation
becomes slower and slower for larger $\beta$. These features indicate the QNMs are endowed with a
{similar property} with the presence of coupling constant $\beta$.

\section{discussion and conclusion}

The main goal of this paper is to study the influence of derivative coupling on the black hole QNMs. The derivative coupling is embodied in the term
$\beta G_{\mu\nu}\nabla^{\mu}\phi\nabla^{\nu}\phi$. Therefore, in order to gain non-trivial results, the background spacetime must not be the solution of vacuum Einstein
equations. So we take into account the RN black holes, the Bardeen black holes, the NGI black holes and the dilaton black holes, respectively. We derive the expression of effective potential with the derivative coupling in the background of general static and spherically symmetric spacetimes. Thus the potential can be applied to arbitrary static spherically symmetric black hole spacetimes. We point that the black holes here are all asymptotically flat in space. For asymptotically de Sitter spacetimes, our potential is still applicable. Since the WKB method can be used for effective potentials which have the form of a potential barrier and take constant values at the event horizon and spatial infinity, one should resort to other method to compute the QNMs of asymptotically de Sitter black holes.
The expressions of effective potential are rather lengthy except for the RN black holes. Thus we do not explicitly present them in the paper
and only make their plotting.

The RN black holes, NGI black holes and Bardeen black holes have two phases. One is for the extreme black holes and the other for the non-extreme black holes.
For the extreme black holes, there is one single horizon. For the non-extreme black holes, there are two horizons. We numerically analyse the potentials, respectively.
The potentials are all asymptotically vanishing at the spatial infinity and the outer event horizon. The height of the potentials is reduced greater and greater with the increasing of coupling constant. The influence of the coupling on the potentials is significant for extreme black holes. For non-extreme black holes, the variation of coupling constant makes negligible influence on the evolution of the potential. The reason for this point is that the coupling term $\beta G_{\mu\nu}\nabla^{\mu}\phi\nabla^{\nu}\phi$ is proportional to the square of the charge.

We calculate the QNMs of black holes for different coupling constants $\beta$ and different overtone numbers $n$. We find that the real part of QNMs is always on the decrease with the increasing $\beta$ and $n$. This signals the oscillation of QNMs become slower and slower with the increasing of $\beta$. On the other hand, the imaginary part is always on the increase with the increasing $\beta$ and $n$. This signals the decay of QNMs become faster and faster with the increasing of $\beta$. {In all, the derivative coupling and the overtone numbers have the similar effects on the quasinormal modes.}

{Finally, what we consider throughout the paper is the QNMs of a massless scalar field. It is found that the QNMs of a {massive} scalar field possess
an important feature: the arbitrary slowly damped quasinormal modes,
called quasi-resonances \cite{Konoplya:2004wg, Konoplya:2006br}. Since the WKB approach can only indicate some trend to quasi-resonances but fails at sufficiently large mass of the field \cite{Konoplya:2017tvu}, one may wonder whether this property will survive with the derivative coupling for the massive scalar field. This is an interesting question and worthy of further study.}

\begin{table*}[h]
\begin{tabular}[b]{cccc}
\hline \hline
 \;\;\;\; $\beta$ \;\;\;\; & \;\;\;\; $\omega\ \ \ (l=1)$\;\;\;\;  & \;\;\;\;  $\omega \ \ \ (l=2)$\;\;\;\;
 & \;\;\;\; $\omega \ \ \ (l=3)$ \;\;\;\; \\ \hline
\\
0.0&\;\;\;\;\;0.3773-0.0959i\;\;\;\;\;  & \;\;\;\;
0.6264-0.0910i\;\;\;\;\;
 & \;\;\;\;\;0.8760-0.0896i\;\;\;\;\;
 \\
0.2&0.3734-0.0993i&0.6192-0.0941i&0.8658-0.0927i
\\
0.4&0.3698-0.1022i&0.6127-0.0969i&0.8566-0.0954i
\\
0.6&0.3665-0.1049i&0.6068-0.0994i&0.8482-0.0978i
\\
0.8&0.3636-0.1072i&0.6014-0.1015i&0.8404-0.1000i
\\
1.0&0.3608-0.1093i&0.5964-0.1035i&0.8332-0.1019i
\\
\hline \hline
\end{tabular}
\begin{center}
\end{center}
\caption{The fundamental ($n=0$) quasinormal frequencies $\omega$ of extreme RN black holes for different $\beta$ and different $l$. We have put $M=1, Q=1, m=0$. }
\end{table*}

\begin{table*}[h]
\begin{tabular}[b]{cccc}
\hline \hline
 \;\;\;\; $\beta$ \;\;\;\; & \;\;\;\; $\omega\ \ \ (l=1)$\;\;\;\;  & \;\;\;\;  $\omega \ \ \ (l=2)$\;\;\;\;
 & \;\;\;\; $\omega \ \ \ (l=3)$ \;\;\;\; \\ \hline
\\
0.0&\;\;\;\;\;0.3639-0.1010i\;\;\;\;\;  & \;\;\;\;
0.6014-0.0964i\;\;\;\;\;
 & \;\;\;\;\;0.8401-0.0951i\;\;\;\;\;
 \\
0.2&0.3511-0.1095i&0.5800-0.1035i&0.8102-0.1019i
\\
0.4&0.3417-0.1157i&0.5637-0.1088i&0.7873-0.1069i
\\
0.6&0.3343-0.1205i&0.5506-0.1128i&0.7687-0.1107i
\\
0.8&0.3282-0.1242i&0.5396-0.1160i&0.7530-0.1137i
\\
1.0&0.3229-0.1273i&0.5301-0.1185i&0.7395-0.1160i
\\
\hline \hline
\end{tabular}
\begin{center}
\end{center}
\caption{The fundamental ($n=0$) quasinormal frequencies $\omega$ of non-extreme RN black holes for different $\beta$ and different $l$. We have put $M=1, Q=0.95, m=0$.}
\end{table*}

\begin{table*}[h]
\begin{tabular}[b]{cccc}
\hline \hline
 \;\;\;\; $\beta$ \;\;\;\; & \;\;\;\; $\omega\ \ \ (l=1)$\;\;\;\;  & \;\;\;\;  $\omega \ \ \ (l=2)$\;\;\;\;
 & \;\;\;\; $\omega \ \ \ (l=3)$ \;\;\;\; \\ \hline
\\
0.0&\;\;\;\;\;0.3337-0.0859i\;\;\;\;\;  & \;\;\;\;
0.5540-0.0807i\;\;\;\;\;
 & \;\;\;\;\;0.7747-0.0792i\;\;\;\;\;
 \\
0.2&0.3309-0.0888i&0.5491-0.0834i&0.7677-0.0819i
\\
0.4&0.3284-0.0914i&0.5446-0.0858i&0.7614-0.0843i
\\
0.6&0.3261-0.0938i&0.5404-0.0881i&0.7555-0.0864i
\\
0.8&0.3240-0.0960i&0.5366-0.0901i&0.7501-0.0884i
\\
1.0&0.3220-0.0980i&0.5331-0.0919i&0.7451-0.0902i
\\
\hline \hline
\end{tabular}
\begin{center}
\end{center}
\caption{The fundamental ($n=0$) quasinormal frequencies $\omega$ of extreme Bardeen black holes for different $\beta$ and different $l$. We have put $M=1, Q=4\sqrt{3}/9, m=0$.}
\end{table*}

\begin{table*}[h]
\begin{tabular}[b]{cccc}
\hline \hline
 \;\;\;\; $\beta$ \;\;\;\; & \;\;\;\; $\omega\ \ \ (l=1)$\;\;\;\;  & \;\;\;\;  $\omega \ \ \ (l=2)$\;\;\;\;
 & \;\;\;\; $\omega \ \ \ (l=3)$ \;\;\;\; \\ \hline
\\
0.0&\;\;\;\;\;0.3325-0.0870i\;\;\;\;\;  & \;\;\;\;
0.5514-0.0819i\;\;\;\;\;
 & \;\;\;\;\;0.7708-0.0805i\;\;\;\;\;
 \\
0.2&0.3298-0.0897i&0.5467-0.0844i&0.7643-0.0829i
\\
0.4&0.3273-0.0921i&0.5424-0.0866i&0.7583-0.0851i
\\
0.6&0.3251-0.0944i&0.5385-0.0887i&0.7527-0.0871i
\\
0.8&0.3230-0.0964i&0.5349-0.0906i&0.7476-0.0889i
\\
1.0&0.3211-0.0984i&0.5315-0.0923i&0.7428-0.0906i
\\
\hline \hline
\end{tabular}
\begin{center}
\end{center}
\caption{The fundamental ($n=0$) quasinormal frequencies $\omega$ of non-extreme Bardeen black holes for different $\beta$ and different $l$. We have put $M=1, Q=0.76, m=0$.}
\end{table*}

\begin{table*}[h]
\begin{tabular}[b]{cccc}
\hline \hline
 \;\;\;\; $\beta$ \;\;\;\; & \;\;\;\; $\omega\ \ \ (l=1)$\;\;\;\;  & \;\;\;\;  $\omega \ \ \ (l=2)$\;\;\;\;
 & \;\;\;\; $\omega \ \ \ (l=3)$ \;\;\;\; \\ \hline
\\
0.0&\;\;\;\;\;0.3649-0.1092i\;\;\;\;\;  & \;\;\;\;
0.5988-0.1051i\;\;\;\;\;& \;\;\;\;\;0.8328-0.1019i\;\;\;\;\;
 \\
0.2&0.3632-0.1096i&0.5958-0.1054i&0.8314-0.1041i
\\
0.4&0.3616-0.1099i&0.5929-0.1057i&0.8273-0.1044i
\\
0.6&0.3600-0.1103i&0.5901-0.1061i&0.8233-0.1048i
\\
0.8&0.3585-0.1107i&0.5874-0.1064i&0.8195-0.1051i
\\
1.0&0.3570-0.1111i&0.5847-0.1068i&0.8157-0.1054i
\\
\hline \hline
\end{tabular}
\begin{center}
\end{center}
\caption{The fundamental ($n=0$) quasinormal frequencies $\omega$ of dilaton black holes for different $\beta$ and different $l$. We have put $M=1, D=0.5, m=0$.}
\end{table*}

\begin{table*}[h]
\begin{tabular}[b]{cccc}
\hline \hline
 \;\;\;\; $\beta$ \;\;\;\; & \;\;\;\; $\omega\ \ \ (l=1)$\;\;\;\;  & \;\;\;\;  $\omega \ \ \ (l=2)$\;\;\;\;
 & \;\;\;\; $\omega \ \ \ (l=3)$ \;\;\;\; \\ \hline
\\
0.0&\;\;\;\;\;0.2806-0.1067i\;\;\;\;\;  & \;\;\;\;
0.4775-0.0965i\;\;\;\;\;
 & \;\;\;\;\;0.6717-0.0939i\;\;\;\;\;
 \\
0.2&0.2797-0.1085i&0.4770-0.0986i&0.6709-0.0960i
\\
0.4&0.2783-0.1101i&0.4764-0.1004i&0.6701-0.0979i
\\
0.6&0.2765-0.1114i&0.4757-0.1020i&0.6694-0.0997i
\\
0.8&0.2744-0.1126i&0.4751-0.1034i&0.6687-0.1013i
\\
1.0&0.2719-0.1137i&0.4744-0.1049i&0.6680-0.1028i
\\
\hline \hline
\end{tabular}
\begin{center}
\end{center}
\caption{The fundamental ($n=0$) quasinormal frequencies $\omega$ of extreme NGI black holes for different $\beta$ and different $l$. We have put $M=1, \theta=0.2758, m=0$.}
\end{table*}

\begin{table*}[h]
\begin{tabular}[b]{cccc}
\hline \hline
 \;\;\;\; $\beta$ \;\;\;\; & \;\;\;\; $\omega\ \ \ (l=1)$\;\;\;\;  & \;\;\;\;  $\omega \ \ \ (l=2)$\;\;\;\;
 & \;\;\;\; $\omega \ \ \ (l=3)$ \;\;\;\; \\ \hline
\\
0.0&\;\;\;\;\;0.2806-0.1067i\;\;\;\;\;  & \;\;\;\;
0.4775-0.0965i\;\;\;\;\;
 & \;\;\;\;\;0.6717-0.0939i\;\;\;\;\;
 \\
0.2&0.2797-0.1085i&0.4770-0.0986i&0.6711-0.0978i
\\
0.4&0.2783-0.1101i&0.4764-0.1004i&0.6704-0.1000i
\\
0.6&0.2765-0.1114i&0.4757-0.1020i&0.6696-0.1015i
\\
0.8&0.2744-0.1126i&0.4750-0.1035i&0.6693-0.1020i
\\
1.0&0.2719-0.1137i&0.4744-0.1049i&0.6690-0.1028i
\\
\hline \hline
\end{tabular}
\begin{center}
\end{center}
\caption{The fundamental ($n=0$) quasinormal frequencies $\omega$ of non-extreme NGI black holes for different $\beta$ and different $l$. We have put $M=1, \theta=0.24, m=0$.}
\end{table*}

\begin{table*}[h]
\centering  % ������
\begin{tabular}{lcccccccc}  % {lccc} ��ʾ����Ԫ�ض��뷽ʽ��left-l,right-r,center-c
\hline
$n$ & $\omega\ (l=1)$ & $\omega\ (l=2)$ & $\omega\ (l=3)$ & $\omega\ (l=4)$ & $\omega\ (l=5)$ & $\omega\ (l=6)$ & $\omega\ (l=7)$ & $\omega\ (l=8)$\\ \hline  % \hline �ڴ������滭һ����

        % \\ ��ʾ���¿�'һ��
0 &0.3608-0.1093i &0.5964-0.1035i &0.8332-0.1018i &1.0704-0.1012i &1.3077-0.1009i &1.5451-0.1007i &1.7826-0.1006i &2.0200-0.1006i\\
1 & &0.5819-0.3168i &0.8210-0.3091i &1.0602-0.3058i &1.2991-0.3040i &1.5377-0.3031i &1.7761-0.3025i &2.0143-0.3021i\\
2 & & & 0.8016-0.5243i &1.0425-0.5157i &1.2834-0.5109i &1.5238-0.5080i &1.7637-0.5062i &2.0031-0.5050i\\
3 & & & &1.0206-0.7318i &1.2626-0.7224i &1.5046-0.7164i &1.7461-0.7125i &1.9871-0.7098i\\
4 & & & & &1.2389-0.9387i &1.4818-0.9288i &1.7245-0.9219i &1.9669-0.9170i\\
5 & & & & & &1.4563-1.1447i &1.6997-1.1344i &1.9432-1.1268i\\
6 & & & & & & &1.6726-1.3496i &1.9166-1.3390i\\
7 & & & & & & & &1.8878-1.5535i\\
\hline
\end{tabular}

\caption{The quansinormal frequencies of extreme RN black holes for different $n$ and different $l$. We have put $M=1, Q=1, m=0, \beta=1$.}
\end{table*}

\begin{table*}[h]
\centering  % ������
\begin{tabular}{lcccccccc}  % {lccc} ��ʾ����Ԫ�ض��뷽ʽ��left-l,right-r,center-c
\hline
$n$ & $\omega\ (l=1)$ & $\omega\ (l=2)$ & $\omega\ (l=3)$ & $\omega\ (l=4)$ & $\omega\ (l=5)$ & $\omega\ (l=6)$ & $\omega\ (l=7)$ & $\omega\ (l=8)$\\ \hline  % \hline �ڴ������滭һ����
0 &0.3511-0.1095i &0.5800-0.1035i &0.8102-0.1019i &1.0408-0.1012i &1.2715-0.1009i &1.5023-0.1007i &1.7332-0.1006i &1.9641-0.1005i\\
1 & &0.5657-0.3171i &0.7980-0.3093i &1.0306-0.3058i &1.2629-0.3041i &1.4949-0.3031i &1.7267-0.3025i &1.9583-0.3021i\\
2 & & &0.7788-0.5247i &1.0130-0.5160i &1.2472-0.5110i &1.4810-0.5081i &1.7142-0.5062i &1.9471-0.5050i\\
3 & & & &0.9914-0.7323i &1.2266-0.7228i &1.4618-0.7167i &1.6967-0.7127i &1.9310-0.7099i\\
4 & & & & &1.2033-0.9392i &1.4391-0.9293i &1.6752-0.9223i &1.9109-0.9173i\\
5 & & & & & &1.4141-1.1453i &1.6506-1.1350i &1.8873-1.1273i\\
6 & & & & & & &1.6239-1.3504i &1.8610-1.3397i\\
7 & & & & & & & &1.8325-1.5542i\\
        % \\ ��ʾ���¿�'һ��
\hline
\end{tabular}
\caption{The quansinormal frequencies of non-extreme RN black holes for different $n$ and different $l$. We have put $M=1, Q=0.95, m=0, \beta=1$.}
\end{table*}

\begin{table*}[h]
\centering  % ������
\begin{tabular}{lcccccccc}  % {lccc} ��ʾ����Ԫ�ض��뷽ʽ��left-l,right-r,center-c
\hline
$n$ & $\omega\ (l=1)$ & $\omega\ (l=2)$ & $\omega\ (l=3)$ & $\omega\ (l=4)$ & $\omega\ (l=5)$ & $\omega\ (l=6)$ & $\omega\ (l=7)$ & $\omega\ (l=8)$\\ \hline  % \hline �ڴ������滭һ����
0 &0.3220-0.0980i &0.5331-0.0919i &0.7451-0.0902i &0.9573-0.0895i &1.1696-0.0892i &1.3819-0.0890i &1.5943-0.0889i &1.8068-0.0888i\\
1 & &0.5177-0.2827i &0.7322-0.2745i &0.9466-0.2709i &1.1606-0.2691i &1.3742-0.2680i &1.5876-0.2674i &1.8008-0.2669i\\
2 & & &0.7120-0.4675i &0.9281-0.4582i &1.1442-0.4530i &1.3597-0.4499i &1.5747-0.4480i &1.7892-0.4466i\\
3 & & & &0.9057-0.6524i &1.1228-0.6423i &1.3399-0.6358i &1.5565-0.6316i &1.7726-0.6286i\\
4 & & & & &1.0987-0.8368i &1.3165-0.8262i &1.5343-0.8187i &1.7517-0.8134\\
5 & & & & & &1.2909-1.0206i &1.5092-1.0095i &1.7276-1.0012i\\
6 & & & & & & &1.4821-1.2034i &1.7009-1.1919i\\
7 & & & & & & & &1.6723-1.3852i\\
        % \\ ��ʾ���¿�'һ��
\hline
\end{tabular}
\caption{The quansinormal frequencies of extreme Bardeen black holes for different $n$ and different $l$. We have put $M=1, Q=4\sqrt{3}/9, m=0, \beta=1$.}
\end{table*}

\begin{table*}[h]
\centering  % ������
\begin{tabular}{lcccccccc}  % {lccc} ��ʾ����Ԫ�ض��뷽ʽ��left-l,right-r,center-c
\hline
$n$ & $\omega\ (l=1)$ & $\omega\ (l=2)$ & $\omega\ (l=3)$ & $\omega\ (l=4)$ & $\omega\ (l=5)$ & $\omega\ (l=6)$ & $\omega\ (l=7)$ & $\omega\ (l=8)$\\ \hline  % \hline �ڴ������滭һ����
0 &0.3211-0.0983i &0.5315-0.9232i &0.7428-0.9061i &0.9543-0.0899i &1.1659-0.0896i &1.3776-0.0894i &1.5893-0.0893i &1.8011-0.0892i\\
1 & &0.5164-0.2838i &0.7301-0.2756i &0.9438-0.2721i &1.1571-0.2702i &1.3700-0.2692i &1.5827-0.2686i &1.7952-0.2681i\\
2 & & &0.7101-0.4692i &0.9255-0.4600i &1.1409-0.4549i &1.3557-0.4518i &1.5699-0.4499i &1.7837-0.4486i\\
3 & & & &0.9034-0.6548i &1.1197-0.6448i &1.3361-0.6384i &1.5520-0.6342i &1.7673-0.6313i\\
4 & & & & &1.0960-0.8399i &1.3130-0.8294i &1.5300-0.8220i &1.7467-0.8167i\\
5 & & & & & &1.2878-1.0243i &1.5052-1.0133i &1.7229-1.0051i\\
6 & & & & & & &1.4786-1.2078i &1.6965-1.1964i\\
7 & & & & & & & &1.6683-1.3902i\\
        % \\ ��ʾ���¿�'һ��
\hline
\end{tabular}
\caption{The quansinormal frequencies of non-extreme Bardeen black holes for different $n$ and different $l$. We have put $M=1, Q=0.76, m=0, \beta=1$.}
\end{table*}

\begin{table*}[h]
\centering  % ������
\begin{tabular}{lcccccccc}  % {lccc} ��ʾ����Ԫ�ض��뷽ʽ��left-l,right-r,center-c
\hline
$n$ & $\omega\ (l=1)$ & $\omega\ (l=2)$ & $\omega\ (l=3)$ & $\omega\ (l=4)$ & $\omega\ (l=5)$ & $\omega\ (l=6)$ & $\omega\ (l=7)$ & $\omega\ (l=8)$\\ \hline  % \hline �ڴ������滭һ����
0 &0.3570-0.1111i &0.5847-0.1068i &0.8157-0.1054i &1.0474-0.1049i &1.2794-0.1046i &1.5116-0.1046i &1.7438-0.1044i &1.9760-0.1043i\\
1 & &0.5729-0.3248i &0.8048-0.3191i &1.0381-0.3165i &1.2714-0.3151i &1.5046-0.3143i &1.7376-0.3137i &1.9705-0.3133i\\
2 & & &0.7871-0.5388i &1.0218-0.5327i &1.2568-0.5288i &1.4916-0.5266i &1.7259-0.5247i &1.9599-0.5235i\\
3 & & & &1.0015-0.7538i &1.2375-0.7467i &1.4739-0.7424i &1.7094-0.7383i &1.9446-0.7355i\\
4 & & & & &1.2156-0.9688i &1.4529-0.9616i &1.6890-0.9545i &1.9253-0.9496i\\
5 & & & & & &1.4298-1.1840i &1.6657-1.1735i &1.9027-1.1659i\\
6 & & & & & & &1.6400-1.3946i &1.8772-1.3842i\\
7 & & & & & & & &1.8494-1.6044i\\
        % \\ ��ʾ���¿�'һ��
\hline
\end{tabular}
\caption{The quansinormal frequencies of dilaton black holes for different $n$ and different $l$. We have put $M=1, D=0.5, m=0, \beta=1$.}
\end{table*}

\begin{table*}[h]
\centering
\begin{tabular}{lcccccccc}
\hline
$n$ & $\omega\ (l=1)$ & $\omega\ (l=2)$ & $\omega\ (l=3)$ & $\omega\ (l=4)$ & $\omega\ (l=5)$ & $\omega\ (l=6)$ & $\omega\ (l=7)$ & $\omega\ (l=8)$\\ \hline
0 &0.2719-0.1138i &0.4744-0.1049i &0.6680-0.1029i &0.8606-0.1020i &1.0530-0.1016i &1.2452-0.1014i &1.4373-0.1014i &1.6294-0.1014i\\
1 & &0.4276-0.3290i &0.6322-0.3149i &0.8315-0.3087i &1.0287-0.3057i &1.2244-0.3042i &1.4192-0.3037i &1.6134-0.3034i\\
2 & & &0.5690-0.5419i &0.7757-0.5231i &0.9800-0.5130i &1.1819-0.5074i &1.3819-0.5046i &1.5802-0.5031i\\
3 & & & &0.6972-0.7486i &0.9074-0.7262i &1.1163-0.7123i &1.3232-0.7045i &1.5276-0.6999i\\
4 & & & & &0.8123-0.9480i &1.0266-0.9217i &1.2407-0.9050i &1.4525-0.8945i\\
5 & & & & & &0.9132-1.1392i &1.1330-1.1094i &1.3524-1.0894i\\
6 & & & & & & &0.9997-1.3221i &1.2250-1.2882i\\
7 & & & & & & & &1.0703-1.4962i\\
\hline
\end{tabular}
\caption{The quansinormal frequencies of extreme NGI black holes for different $n$ and different $l$. We have put $M=1, \theta=0.2758, m=0, \beta=1$.}
\end{table*}

\begin{table*}[h]
\centering
\begin{tabular}{lcccccccc}
\hline
$n$ & $\omega\ (l=1)$ & $\omega\ (l=2)$ & $\omega\ (l=3)$ & $\omega\ (l=4)$ & $\omega\ (l=5)$ & $\omega\ (l=6)$ & $\omega\ (l=7)$ & $\omega\ (l=8)$\\ \hline
0 &0.2719-0.1138i &0.4744-0.1049i &0.6680-0.1028i &0.8606-0.1020i &1.0530-0.1016i &1.2452-0.1014i &1.4373-0.1013i &1.6294-0.1013i\\
1 & &0.4275-0.3289i &0.6321-0.3147i &0.8315-0.3087i &1.0287-0.3058i &1.2244-0.3042i &1.4192-0.3035i &1.6133-0.3030i\\
2 & & &0.5689-0.5416i &0.7757-0.5231i &0.9801-0.5132i &1.1819-0.5074i &1.3818-0.5043i &1.5800-0.5024i\\
3 & & & &0.6973-0.7486i &0.9075-0.7263i &1.1163-0.7123i &1.3229-0.7041i &1.5271-0.6989i\\
4 & & & & &0.8125-0.9482i &1.0266-0.9218i &1.2404-0.9046i &1.4518-0.8934i\\
5 & & & & & &1.0266-0.9218i &1.1325-1.1090i &1.3514-1.0882i\\
6 & & & & & & &0.9991-1.3217i &1.2237-1.2870i\\
7 & & & & & & & &1.0685-1.4950i\\
\hline
\end{tabular}
\caption{The quansinormal frequencies of non-extreme NGI black holes for different $n$ and different $l$. We have put $M=1, \theta=0.24, m=0, \beta=1$.}
\end{table*}

\section*{Acknowledgments}
We are very grateful to the referees for the expert suggestions which have improved the paper significantly. Shuang Yu is very grateful to Xiao Guo, Chen Wang, Weiyang Wang, Xing Wang and Shifan Zuo for their great help
during this research. This work is partially supported by China Program of International ST Cooperation 2016YFE0100300,
the Strategic Priority Research Program ``Multi-wavelength Gravitational Wave Universe'' of the CAS,
Grant No. XDB23040100, the Joint Research Fund in Astronomy (U1631118), and the NSFC
under grants 11473044, 11633004 and the Project of CAS, QYZDJ-SSW-SLH017.

\newcommand\ARNPS[3]{~Ann. Rev. Nucl. Part. Sci.{\bf ~#1}, #2~ (#3)}
\newcommand\AL[3]{~Astron. Lett.{\bf ~#1}, #2~ (#3)}
\newcommand\AP[3]{~Astropart. Phys.{\bf ~#1}, #2~ (#3)}
\newcommand\AJ[3]{~Astron. J.{\bf ~#1}, #2~(#3)}
\newcommand\APJ[3]{~Astrophys. J.{\bf ~#1}, #2~ (#3)}
\newcommand\APJL[3]{~Astrophys. J. Lett. {\bf ~#1}, L#2~(#3)}
\newcommand\APJS[3]{~Astrophys. J. Suppl. Ser.{\bf ~#1}, #2~(#3)}
\newcommand\JHEP[3]{~JHEP.{\bf ~#1}, #2~(#3)}
\newcommand\JMP[3]{~J. Math. Phys. {\bf ~#1}, #2~(#3)}
\newcommand\JCAP[3]{~JCAP {\bf ~#1}, #2~ (#3)}
\newcommand\LRR[3]{~Living Rev. Relativity. {\bf ~#1}, #2~ (#3)}
\newcommand\MNRAS[3]{~Mon. Not. R. Astron. Soc.{\bf ~#1}, #2~(#3)}
\newcommand\MNRASL[3]{~Mon. Not. R. Astron. Soc.{\bf ~#1}, L#2~(#3)}
\newcommand\NPB[3]{~Nucl. Phys. B{\bf ~#1}, #2~(#3)}
\newcommand\CMP[3]{~Comm. Math. Phys.{\bf ~#1}, #2~(#3)}
\newcommand\CQG[3]{~Class. Quantum Grav.{\bf ~#1}, #2~(#3)}
\newcommand\PLB[3]{~Phys. Lett. B{\bf ~#1}, #2~(#3)}
\newcommand\PLA[3]{~Phys. Lett. A{\bf ~#1}, #2~(#3)}
\newcommand\PRL[3]{~Phys. Rev. Lett.{\bf ~#1}, #2~(#3)}
\newcommand\PR[3]{~Phys. Rep.{\bf ~#1}, #2~(#3)}
\newcommand\PRd[3]{~Phys. Rev.{\bf ~#1}, #2~(#3)}
\newcommand\PRD[3]{~Phys. Rev. D{\bf ~#1}, #2~(#3)}
\newcommand\RMP[3]{~Rev. Mod. Phys.{\bf ~#1}, #2~(#3)}
\newcommand\SJNP[3]{~Sov. J. Nucl. Phys.{\bf ~#1}, #2~(#3)}
\newcommand\ZPC[3]{~Z. Phys. C{\bf ~#1}, #2~(#3)}
\newcommand\IJGMP[3]{~Int. J. Geom. Meth. Mod. Phys.{\bf ~#1}, #2~(#3)}
\newcommand\IJMPD[3]{~Int. J. Mod. Phys. D{\bf ~#1}, #2~(#3)}
\newcommand\GRG[3]{~Gen. Rel. Grav.{\bf ~#1}, #2~(#3)}
\newcommand\EPJC[3]{~Eur. Phys. J. C{\bf ~#1}, #2~(#3)}
\newcommand\PRSLA[3]{~Proc. Roy. Soc. Lond. A {\bf ~#1}, #2~(#3)}
\newcommand\AHEP[3]{~Adv. High Energy Phys.{\bf ~#1}, #2~(#3)}
\newcommand\Pramana[3]{~Pramana.{\bf ~#1}, #2~(#3)}
\newcommand\PTP[3]{~Prog. Theor. Phys{\bf ~#1}, #2~(#3)}
\newcommand\APPS[3]{~Acta Phys. Polon. Supp.{\bf ~#1}, #2~(#3)}
\newcommand\ANP[3]{~Annals Phys.{\bf ~#1}, #2~(#3)}

\end{document}